\documentclass[notitlepage]{article}

\usepackage[paper=letterpaper,hmargin=2.5cm,top=2.5cm,bottom=2.5cm,bindingoffset=0cm,twoside,hoffset=0pt,marginparsep=0pt,marginparwidth=0pt]{geometry}
\usepackage{feynmp-auto}
\usepackage{pdfpages}
\usepackage{graphicx}
\usepackage{dcolumn} % Align table columns on decimal point
\usepackage{bm} % bold math
\usepackage{hyperref}
\usepackage{mathrsfs}
\usepackage{amsmath}
\usepackage{amssymb} % contains \Box
\usepackage{xcolor}
\usepackage{appendix}
\usepackage{mathtools} %  \overbrace{}^ y \underbrace{}_

\usepackage[utf8]{inputenc}
\usepackage[T1]{fontenc}

\title{The $b$ quark fragmentation fractions at LHCb and meson decays with heavy quark spectators
}

\author{
C.O. Dib$^1$\footnote{E-mail: claudio.dib@usm.cl},
C. S. Kim$^2$\footnote{E-mail: cskim@yonsei.ac.kr},
N.A. Neill$^3$\footnote{E-mail: naneill@outlook.com}\\
\small
$^1$ \textit{Dept.\,of Physics and CCTVal, Federico Santa Maria Technical University, Valparaíso, Chile}\\
\small
\textit{$^2$ Dept.\,of Physics and IPAP, Yonsei University, Seoul 120-749, Korea}
\\
\small
$^3$ \textit{Dept.\,of Electrical-Electronic Engineering, University of Tarapacá, Arica, Chile}
\normalsize
\vspace{-4mm}
}
\date{}

\begin{document}
\maketitle

\begin{abstract}
We study the current estimates of $B_c\to B_s \pi$ to extract the fragmentation fraction $f_c/f_s$ at the LHCb. A rather robust estimate of $Br(B_c\to B_s \pi)$ based on factorization and lattice results for the form factor gives $f_c/f_s \sim 0.056$ with a 16\% error. We also revisit the extraction of $f_s/f_d$ using $B\to D\pi$ instead of the theoretical cleaner but more suppressed channel $B\to DK$. 
We also find a tension on the predictions of $Br(B_c\to J/\psi \pi)$ and $Br(B_c\to B_s\pi)$ considering the measurements of these modes at LHCb, and find that, within a 23\% uncertainty, only the lower end of the current prediction range $Br(B_c\to J/\psi)\sim 0.4\% - 1.7\%$ would be consistent with the LHCb measurements.
\end{abstract}

\noindent

\section{Introduction}
The heavy meson decay $B_c\to B_s \pi$ is a remarkable case of a decay of a heavy meson where the heavy quark is a spectator of the weak interaction process, in contrast to the much studied transitions such as $B\to D$, where it is the heavy quark that decays and the light degrees of freedom act as spectators. The latter case has been extensively studied, giving birth to the systematic treatment called heavy quark effective theory (HQET), which is a expansion in operators with coefficients in powers of $\Lambda_{QCD}/m_b$ and $\Lambda_{QCD}/m_c$ \cite{Eichten:1989zv, Georgi:1990um,Neubert:1993mb}.
As such, in the infinite mass limit, the results are relatively simple \cite{Nussinov:1986hw, Shifman:1987rj,Isgur:1989vq}. For example, for a transition $B\to D \ell \nu$  at the kinematic point where $B$ and $D$ are at relative rest (the so-called zero recoil point), the wave functions of the two heavy mesons should be the same and the overlap of the form factors should be equivalent to their normalization (in the appropriate normalization, this means the form factor should be unity). In contrast, here we are interested in transitions such as $B_c\to B_s$, where the heavy (or better said the \emph{heaviest}) quark is the spectator and the weak transition occurs in the lighter degrees of freedom. Now the wavefunctions will not coincide, not even at zero recoil, so there is no \emph{a priori} known value for the form factor at any kinematic point. These decays, where the heavy quark is a spectator, therefore offer new information for  hadron structure and bound states under the strong interactions. To have an intuitive picture let us consider the heavy meson as a bound state of two constituent quarks. The Bohr radius of the wavefunction goes as the inverse of the reduced mass. In decays such as $B\to D \ell\nu$,  in the limit $m_b, m_c\to \infty$ the reduced mass is the same in both mesons, so one expects that the form factor at zero recoil, which is the overlap of the $B$ and $D$ wavefunctions, should be the same as their normalization (i.e. the form factor goes to a known value, usually taken as unity). However in $B_c\to B_s$ transitions the $B_c$ and $B_s$ reduced masses and Bohr radii are different, so the form factor at zero recoil is not a fixed norm but depends on the bound state dynamics. 
The first measurement (and so far the only one) of a $B_c \to B_s$ transition was done by LHCb \cite{LHCb:2013xlg}
in the two-body decay $B_c^+\to B_s \pi^+$, however the result on this branching ratio comes combined with the ratio of the fragmentation fractions of the $b$ quarks produced in the $pp$ collisions into the different $B$ mesons, in this case $\sigma(B_c)/\sigma(B_s) \equiv f_c/f_s$. 
It is therefore necessary to have an independent estimate of either the branching ratio or of $f_c/f_s$, in order to extract a value for the other. 
In Ref. \cite{LHCb:2013xlg} the authors exemplify the use of independent $f_c/f_s$ estimates to determine the $B_c\to B_s\pi$ branching ratio.  
Here we will take the opposite avenue, namely to use 
the best theoretical estimate we find for the branching ratio, which is based on current lattice results for the form factors \cite{Cooper:2020wnj},
 in order to extract the value of $f_c/f_s$ from the LHCb measurement.  Using this estimate for $f_c/f_s$ we then obtain estimates for other $B_c$ branching ratios such as $B( B_c^+ \to J/\psi \pi^+)$, which in Ref.\cite{LHCb:2013xlg} was used as an input. 

Concerning the $b$ fragmentation fractions at LHCb, another important ratio is $f_s/f_d$. This ratio was estimated earlier at LHCb, using the decays $B_s\to D_s^- \pi^+$ 
and $B^0\to D^-K^+$ \cite{LHCb:2013vfg}, and more recently doing a global fit using several other decays \cite{LHCb:2021qbv}, thus reducing the uncertainty. In the former study, the reason for using $B\to D K$ instead of the more numerous $B\to D\pi$ was that its theoretical calculation contains an additional amplitude that is more difficult to estimate. Here we revisit these estimates using updated $B$ meson branching ratios as well as more recent lattice results for the relevant form factors. 
In Section II we address the analysis of $f_c/f_s$ and $B_c\to B_s$ transitions. In Section III we revisit the estimates and uncertainties of $f_s/f_d$ and the non-factorizable amplitude of $B\to D\pi$. In Section IV we revisit $Br(B_c\to J/\psi\, \pi^+)$ and the extraction of $f_c/f_d$.  Our conclusions are in Section V.

\section{$B_c\to B_s$ transitions and $f_c/f_s$}

The decay $B_c\to B_s\pi$ was observed at LHCb \cite{LHCb:2013xlg} and its branching ratio was determined up to the factor $f_c/f_s$:
\begin{equation}
\frac{f_c}{f_s} \times Br(B_c^+\to B_s \pi^+)  = (2.37 \pm 0.31 \pm 0.11 {}^{+0.17}_{-0.13})\times 10^{-3} .
\label{BcBs}
\end{equation}
Here $f_c$ and $f_s$ are the fragmentation fractions, i.e.\, the probabilities that a $b$ quark produced in the $pp$ collisions hadronizes into a $B_c$ or a $B_s$ meson, respectively, and $Br$ stands for \emph{branching ratio}. 
The ratio $f_c/f_s$ appears because one does not know \emph{a priori} the number of $B_c$ produced in the collisions, and so the count of $B_c \to B_s\pi$ events is given relative to the number of $B_s$ produced. 
Since in that analysis the $B_s$ were identified only by their decays $B_s\to D_s\pi$ and $B_s\to J/\psi \,\phi$, then $B_c\to B_s\pi$  are identified by the events $(D_s\pi)\pi$ and $(J/\psi\ \phi)\pi$, where the pair reconstructs the ${B_s}$ mass and the pair plus the bachelor pion reconstructs the $B_c$ mass. 
One can understand Eq.\eqref{BcBs} by first denoting $B_s\to D_s \pi$ and $B_s\to J/\psi \phi$ generically as $B_s\to X$; then the number of observed events $B_c\to X \pi$, where $X$ reconstructs the $B_s$ mass, can be expressed as:
\[
n(B_c\to X\pi) = N_b\cdot f_c \cdot Br(B_c\to B_s \pi) \cdot Br(B_s\to X)
\]
where 
$N_b$ is the number of $b$ quarks produced in the $pp$ collisions and
$N_b\cdot f_c$ --an unknown--  is the total number of $B_c$ produced. In the same way, the total number of observed $B_s\to X$ events (regardless whether $B_s$ comes from a $B_c$ decay or not) is:
\[
n(B_s\to X) = N_b\cdot f_s \cdot Br(B_s\to X).
\]
Dividing these two expressions one obtains Eq.\eqref{BcBs}, where the ratio of counts $n(B_c\to X \pi)/n(B_s \to X)$ is the  result on the right hand side, measured by LHCb  \cite{LHCb:2013xlg}.

In order to separate  $Br(B_c\to B_s\pi)$ from $f_c/f_s$ in Eq. \eqref{BcBs}, one needs an independent input. In Ref \cite{LHCb:2013xlg} this was done by separately estimating
$f_c/f_s$, expressing it as:
\[
\frac{f_c}{f_s} = \frac{f_c/f_d}  {f_s/f_d},
\]
 where the denominator was previously given in 
 Ref.\cite{LHCb:2013vfg} as $f_s/f_d = 0.256 \pm 0.020$, and the numerator $f_c/f_d$ was extracted from a previous LHCb result \cite{LHCb:2012ihf}:
\begin{equation}
\frac{f_c}{f_d} \times \frac{Br(B_c\to J/\psi \,\pi)}{Br(B\to J/\psi\, K)} = 
\left( 0.68 \pm 0.10\pm 0.03 \pm 0.05
\right) \% .
\label{fcfd}    
\end{equation}
While $Br(B\to J/\psi\,K) = \left( 1.016 \pm 0.033\right) \times 10^{-3}$ has been measured \cite{ParticleDataGroup:2012pjm},  for 
$
Br(B_c\to J/\psi\,\pi) \sim 0.06\% \to 0.18\%
$
these are  only theoretical estimates \cite{Ivanov:2006ni}. The latter clearly brought the largest uncertainty, 
while the next largest uncertainties are  $\pm 17\%$ in Eq.\eqref{fcfd}, $\pm 15\%$ in Eq.\eqref{BcBs} and
$\pm 7.8\%$ in $f_s/f_d$ according to Ref. \cite{LHCb:2013vfg}.
Consequently the extracted $Br(B_c \to B_s \pi)$ 
from this method 
reproduces the large uncertainty present in $Br(B_c\to J/\psi\, \pi)$:
\begin{equation}
Br(B_c\to B_s\pi) \sim 5\% \to 15\%,
\label{BcBsLHCb}
\end{equation}
and similar uncertainties were then  obtained for $f_c/f_s \sim 1.6\% \to 4.6\%$ and $f_c/f_d \sim 0.4\% \to 1.2\%$.

Here we take the opposite avenue, namely to start from current lattice estimates for the relevant form factor of $Br(B_c\to B_s\pi)$ \cite{Cooper:2020wnj} to derive the value of $f_c/f_s$ from Eq.\eqref{BcBs}.

The expression for the amplitude of $B^+_c \to B^0_s\pi^+$, assuming factorization, is~\cite{AbdEl-Hady:1999jux},
\begin{equation}
    \mathcal M (B_c^+\to B_s \pi^+) = \frac{G_F}{\sqrt{2}} V_{cs}V^*_{ud}\ a_1 \, \left<\pi^+|(\bar u d)|0\right>
    \left<B_s|(\bar s c)|B_c\right>,\label{eq:MBctoBspi}
\end{equation}
where $(\bar u d)$ and ($\bar s c)$ denote $V-A$ color singlet currents.
Within this assumption the matrix element of the 4-quark operator is separated into a product of current matrix elements.  Non-factorizable contributions, where gluons are exchanged between these currents, are neglected (for more details see Section III). 

Moreover,  $a_1$ is the corresponding combination of Wilson coefficients at the renormalization scale of $B$ mesons, $\mu \sim m_b$ \cite{Buras:1994ij}. In Ref. \cite{Buras:1994ij}, a value 
$a_1 \simeq 1.02 \pm 0.01 $ is estimated, but 
one can find other estimates as well, which depend also on the flavor transition. 
Larger values at the $B$ scale have also been used, e.g. $a_1 = 1.14$ \cite{Ivanov:2006ni,Ebert:2003cn,Kiselev:2002vz}, or even $a_1 = 1.26$ \cite{Kiselev:2000pp}. Several authors have used a value $a_1 \simeq 1.2$ for transitions $c\to s$ in beauty meson decays such as $B_c \to B_s \pi$, and a value $a_1 \simeq 1.14$ for transitions $b\to c$ such as $B_c \to J/\psi\pi$ 
\cite{Kiselev:2002vz,Kiselev:2000pp, Ebert:2003cn, Ebert:2003wc, AbdEl-Hady:1999jux, Chang:1992pt, Anisimov:1998xv,Liu:1997hr, Ivanov:2006ni}.
Since $a_1$ is a combination of Wilson coefficients that integrate QCD corrections from the electroweak scale $M_W$ down to $\mu\sim m_b$, its estimates can be done systematically. In a detailed work that includes next-to-leading-order (NLO) calculations 
the authors of Ref. 
\cite{Beneke:2000ry}
find $a_1 \simeq 1.05$ with a $2\%$ uncertainty by varying 
the scale $\mu \sim m_b/2 \to 2 m_b$. A more recent result at NNLO 
can be found in Ref. 
\cite{Huber:2016xod}
where $a_1=1.070_{-0.013}^{+0.010}$ is obtained for $\overline B_d\to D^+K^-$ and very similar results for other final state mesons; here the quoted errors include the uncertainties from the variation of the scale $\mu$ of the Wilson coefficients, quark masses, Gegenbauer moments of the light meson wave functions, and the strong coupling  $\alpha_s(m_Z)$. \cite{Huber:2016xod}
In order to separate the uncertainties in $B$ meson decays coming from $a_1$, wherever possible  we will cite the results on theoretical branching ratios  with $a_1$ exhibited as an explicit factor (see Tables I and II).

In Eq. \eqref{eq:MBctoBspi} the first matrix element involves the pion decay constant $f_\pi$ and the pion momentum $q_\mu$:
\begin{equation}
    \left<\pi^+|(\bar u d)|0\right> = if_\pi q_\mu .
\end{equation}
The second matrix element involves two form factors \cite{Cooper:2020wnj}:
\begin{equation}
\left<B_s(p_2)|(\bar s c)|B_c(p_1)\right>    
= f_0(q^2) \left(
\frac{m_{B_c}^2 - m_{B_s}^2}{q^2} q^\mu \right)
+ f_+(q^2) \left( p_1^\mu + p_2^\mu - \frac{m_{B_c}^2 - m_{B_s}^2}{q^2} q^\mu
\right) .\label{eq:ffBstoBc}
\end{equation}
The term with $f_+(q^2)$ vanishes when contracted with $q^\mu$, so we only need the longitudinal form factor $f_0(q^2)$ at $q^2 = m_\pi ^2$:
\begin{equation}
    \mathcal M (B_c^+\to B_s \pi^+) = i \frac{G_F}{\sqrt{2}} V_{cs}V^*_{ud}\  a_1\  f_\pi\  f_0(m_\pi^2 ) 
    \left( m_{B_c}^2 - m_{B_s}^2
    \right) .
\label{AmplitudeBcBspi}    
\end{equation}
The value of the form factor according to lattice fits in the physical-continuum limit is \cite{Cooper:2020wnj}:
\begin{equation}
    f_0(m_\pi^2) = 0.624 \pm 0.011.\label{eq:f02}
\end{equation}
It is now straightforward to obtain the decay rate:
\begin{equation}
\Gamma ( B_c^+ \to B_s \pi^+) = \frac{p_{CM}}{8\pi m_{B_c}^2}  |{\cal M}|^2 ,
\label{decayrate}
\end{equation}
where $p_{CM} =\sqrt{(m_{B_c}^2 - m_{B_s}^2 + m_\pi^2)^2 - 4 m_{B_c}^2 m_\pi^2}/(2 m_{B_c}) \simeq 831.84 \,\mbox{MeV}$ is the 3-momentum of the final state mesons in the CM frame.
Here we use $m_{B_c} = 6,274.47\pm 0.32$ MeV, $m_{B_s} = 5,366.92\pm 0.10$ MeV, $m_\pi = 139.57039$ MeV, $V_{ud} =  0.97373 \pm 0.00031$, $V_{cs}= 0.975 \pm 0.006$, $f_\pi = 130.2 \pm 1.2$ MeV.
To obtain the branching ratio
we use the experimental value of the width $\Gamma(B_c) \simeq (1.291 \pm 0.023) \times 10^{-3}\,\mbox{eV}$.

The prediction for the branching ratio using the data above is then:
\begin{equation}
Br(B_c^+\to B_s \pi^+) = (2.942 \pm 0.104)\times 10^{-2}\  a_1^2.
\label{branchBcBs}
\end{equation}
The uncertainty in this branching ratio is dominated by $f_0^2(m_\pi^2)$ [see Eq.\eqref{eq:f02}] which is near $3.5\%$.
We have left out the factor $a_1^2$ as explained above, as it is usually done in the literature. 
Besides $a_1$, there is the uncertainty of neglecting non-factorizable contributions, but
these are not expected to be large for a decay such as $B_c\to B_s\pi^+$, a \emph{Class 1} decay \cite{Buras:1994ij,AbdEl-Hady:1999jux}, where only a charged meson with large momentum --in this case $\pi^+$-- is generated directly from the color-singlet current.

%%%%%% TABLE %%%%%%%%%%%%%%%%%%%%%%%%

\renewcommand{\arraystretch}{1.1}%

\begin{table}[h]
\footnotesize
    \centering
\begin{tabular}{l|c}
Model
  & $Br(B_c\to B_s \pi$)\\
\hline\hline
KKL \cite{Kiselev:2002vz,Kiselev:2000pp}
  &  $1.14\times 10^{-1}\,a_1^2$\\
EFG \cite{Ebert:2003wc}
  & $1.75\times 10^{-2}\,a_1^2$\\
AMV \cite{AbdEl-Hady:1999jux}
  & $1.10\times 10^{-2}\,a_1^2$\\
CC \cite{Chang:1992pt}
  & $3.55\times 10^{-2}\,a_1^2$
  \\
CD \cite{Colangelo:1999zn}
  & $2.14\times 10^{-2}\,a_1^2$
  \\
AKNT \cite{Anisimov:1998xv}
  & $3.33\times 10^{-2}\,a_1^2$
  \\
LC \cite{Liu:1997hr} 
  & $4.55\times 10^{-2}\,a_1^2$ 
  \\
IKS \cite{Ivanov:2006ni}
  & $2.71\times 10^{-2}\, a_1^2$
  \\
\hline
Ours from HPQCD \cite{Cooper:2020wnj} 
  & $2.94\times 10^{-2}\,a_1^2$
  \\
\hline
\hline
\end{tabular}    
    \caption{Predictions for the branching ratio $Br(B_c\to B_s\pi)$ from different authors. The values listed are the partial widths extracted from their respective cited article divided by the full width also cited therein. 
     The last line is our calculation based on the lattice result for the form factor as explained in the text.
    }
    \label{tab:brs-models}
\normalsize
\end{table}

\renewcommand{\arraystretch}{1.0}%

%%%%%%%% END TABLE %%%%%%

Our result in Eq.\eqref{branchBcBs} for $Br(B_c^+\to B_s \pi^+)$ is quite similar to several of the other theoretical predictions shown in Table \ref{tab:brs-models}, and almost all of them are lower than the LHCb estimate of Eq.\eqref{BcBsLHCb} based on the theoretical predictions for $Br(B_c^+\to J/\psi\, \pi^+)$. Therefore the LHCb results shown in Eqs. \eqref{BcBs} and \eqref{fcfd} reveal a tension in the theoretical predictions of $Br(B_c^+\to J/\psi\, \pi^+)$ and 
$Br(B_c^+\to B_s \pi^+)$. If we trust the 
result of  Eq.\eqref{branchBcBs} and most predictions in Table 1 for $Br(B_c^+\to B_s \pi^+)$, we conclude that the actual value of $Br(B_c^+\to J/\psi\, \pi^+)$ should be smaller than the lowest estimate $\sim 0.06\%$ considered in Ref. \cite{LHCb:2013xlg}.

 We will review the branching ratio $Br(B_c^+\to J/\psi\, \pi^+)$ in Section IV. In the present Section we focus on obtaining $f_c/f_s$ based on the theoretical estimate for $Br(B_c^+\to B_s \pi^+)$ in Eq.\eqref{branchBcBs}, and the LHCb measurement \cite{LHCb:2013xlg} shown in Eq.\eqref{BcBs}. A straightforward calculation gives the result:
\begin{equation}
\frac{f_c}{f_s} = \left( 8.06 \pm 1.22\pm 0.28\right) \times 10^{-2}\,a_1^{-2},
\label{fcfsnew}    
\end{equation}
where $a_1 \simeq 1.2$ is the accepted value for $c\to s$ transitions \cite{Kiselev:2002vz,Kiselev:2000pp, Ebert:2003cn, Ebert:2003wc, AbdEl-Hady:1999jux, Chang:1992pt, Anisimov:1998xv,Liu:1997hr, Ivanov:2006ni}. The first uncertainty is from the LHCb measurement \cite{LHCb:2013xlg} shown in Eq.~(\ref{BcBs}), which is about $\sim 15\%$, and the second uncertainty is from the lattice estimate of the form factor \cite{Cooper:2020wnj}, which is near $3.5\%$. Clearly the first uncertainty dominates.
In the next Section we examine $f_s/f_d$.

\section{On $f_s/f_d$ and $B_s \to D_s$ transitions}

\begin{figure}[h]
    \centering
    \includegraphics[scale=0.5]{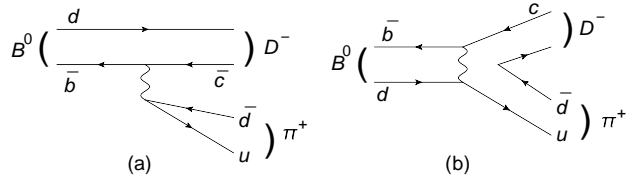}
    \caption{Tree level diagrams for $B^0\to D^-\pi^+$: (a) color-allowed, dominant diagram; (b) color-suppressed, internal $W$-exchange diagram. }
    \label{fig:class3}
\end{figure}

Concerning the determination of the ratio of fragmentation fractions $f_s/f_d$, LHCb has first used $B_s\to D_s\pi$ vs. $B\to D K$  in $pp$ collisions at 7 TeV \cite{LHCb:2013vfg}, and lately \cite{LHCb:2021qbv} used several non-leptonic and semileptonic decays of $B_s$ and $B$ at 7, 8 and 13 TeV \cite{LHCb:2011leg,LHCb:2019fns,LHCb:2013vfg,LHCb:2020zae,LHCb:2019lsv}. In their first study, a ratio of events analogous to Eq.\eqref{fcfd} was obtained:
\begin{equation}
 \frac{f_s}{f_d} \times \frac{Br(B_s \to D_s^-\pi^+)}{Br(B^0\to D^-K^+)} \ =\  \frac{n_{cor}(B_s\to D_s\pi)}{n_{cor}(B^0\to DK)} ,
 \label{fsd}
\end{equation}
where we have denoted $n_{cor}(B_s \to D_s\pi)$  the efficiency-corrected number of $B_s\to D_s\pi$ events and a similar definition for $n_{cor}(B^0\to DK)$. To extract $f_s/f_d$ from Eq.\eqref{fsd} one needs independent estimates of the branching ratios. For that purpose, Ref. \cite{LHCb:2013vfg} uses the theoretical expressions based on factorization, where the main theoretical uncertainties are (i) the ratio of the form factors in $\langle D_s|J^\mu|B_s\rangle$ and $\langle D^-|J^\mu|B^0\rangle$,  and (ii) the neglect of nonfactorizable contributions. Moreover, they use $B^0\to D^-K^+$ instead of the more abundant mode $B^0 \to D^-\pi^+$ precisely because the latter mode has an additional diagram with an internal $W$ exchange, as shown in Fig.\ref{fig:class3}, for which factorization may not be a good approximation \cite{Fleischer:2010ay}.

Here we consider two other options in this analysis: on the one hand, today we have good experimental data for $B^0\to D^-\pi^+$ and $B^0\to D^- K^+$ so we may try to use the $B^0 \to D^-\pi^+$ data instead of the theoretical estimates to extract $f_s/f_d$ from the result in Eq.\eqref{fsd}, provided we have good theoretical estimates for $Br(B_s\to D_s^-\pi^+)$ on the numerator. On the other hand, there are theoretical estimates for the subdominant diagram of $B^0\to D^-\pi^+$ (see Fig.\ref{fig:class3}), so we can verify the uncertainty they bring to the determination of $f_s/f_d$.

\begin{figure}[h]
    \centering
    \includegraphics[scale=0.8]{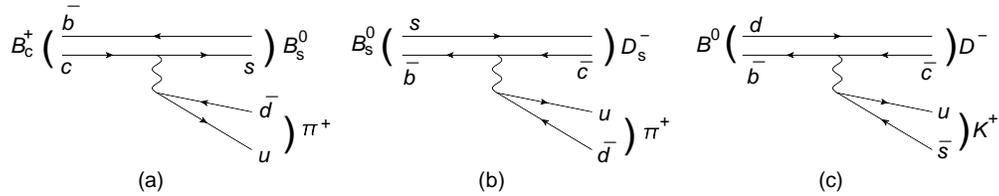}
    \caption{Tree level diagrams for (a) $B_c^+\to B_s\pi^+$, (b) $B_s\to D_s^- \pi^+$ and (c) $B^0\to D^-K^+$. All of them are \emph{color allowed} diagrams in the factorization hypothesis.}
    \label{fig:class1}
\end{figure}

The tree level amplitude for $B_s\to D^-_s\pi^+$ has a similar diagram as $B_c\to B_s\pi^+$ (see Fig.\ref{fig:class1})  and therefore a similar expression as  Eq.\eqref{AmplitudeBcBspi}:
\begin{equation}
    \mathcal M (B_s\to D_s^- \pi^+) = i \frac{G_F}{\sqrt{2}} V_{cb}^*V_{ud}\  a_1\  f_\pi\  f_0^{(s)}(m_\pi^2 ) 
    \left( m_{B_s}^2 - m_{D_s}^2
    \right) ,
\label{AmplitudeBsDspi}    
\end{equation}
but the form factor $f_0^{(s)}(q^2)$ corresponds to the current matrix element $\langle D_s |(\bar b c)|B_s\rangle$.
Similarly the tree level amplitude for $B^0\to D^-K^+$ is:
\begin{equation}
    \mathcal M (B^0\to D^- K^+) = i \frac{G_F}{\sqrt{2}} V_{cb}^*V_{us}\  a_1\  f_K\  f_0^{(d)}(m_K^2 ) 
    \left( m_{B}^2 - m_{D}^2
    \right) ,
\label{AmplitudeBDK}    
\end{equation}
with the form factor $f^{(d)}_0(q^2)$ corresponding to 
$\langle D |(\bar b c)|B\rangle$.

Let us make a brief parenthesis here to review the significance of non factorizable contributions. All of these two-body nonleptonic weak decays can be treated by means of a low-energy  effective weak interaction hamiltonian \cite{Buras:1994ij}, which at leading order contains two 4-quark operators, $O_1$ and $O_2$, both products of two $V-A$ color singlet currents, the former a product of charged currents and the latter of neutral currents. 
As an example, the effective interaction hamiltonian for the decay $B_s\to D_s^-\pi^+$ is:
\[
{\cal H}_{eff} = \frac{G_F}{\sqrt{2}}V_{ud} V^*_{cb}  \Big\{
C_1(\mu)\  O_1 + C_2(\mu)\  O_2 \Big\}, 
\]
where 
$O_1 \equiv  ( \bar u d) (\bar b c)$ and $O_2 \equiv  (\bar u c) (\bar b d)$ (factors in parenthesis represent color-singlet $V-A$ currents). $C_i(\mu)$ are Wilson coefficients at the renormalization scale $\mu \sim m_b$. 
The matrix elements of these 4-quark operators, $\langle D_s^-\pi^+|O_1|B_s\rangle$ and $\langle D_s^-\pi^+|O_2|B_s\rangle$, are difficult to calculate because of the strong interactions.
Factorization is the assumption that these matrix elements can be separated as the product of current matrix elements. Looking at the flavor content of the mesons, the first matrix element could be approximated as:
\begin{equation}
\langle D_s^-\pi^+|O_1|B_s\rangle  \to  \langle\pi^+|( \bar u d)|0\rangle \ 
\langle D_s^-|(\bar b c) |B_s\rangle   \ \equiv \langle O_1\rangle_F.
\label{O1F}
\end{equation}
Since the currents are color singlets, this is expected to be a good approximation.  Instead, the matrix element of $O_2$ requires a Fierz rearrangement, causing a color-suppressed factorized term and a non factorizable one:
\begin{align}
\langle D_s^-\pi^+|O_2|B_s\rangle\  \to
&
\frac{1}{N_c} \langle\pi^+|( \bar u d)|0\rangle \ \langle D_s^-|(\bar b c) |B_s\rangle   + 2 \ \langle D_s^- \pi^+|( \bar u\,  t^a \, d)(\bar b \, t^a\,  c) |B_s\rangle  
\nonumber
\\& \qquad
= \frac{1}{N_c} \langle O_1\rangle_F \quad + \quad nonfact.
\label{O1Fsup}
\end{align}
Notice that, in this case, the factorized matrix element is still $\langle O_1\rangle_F$.
This is the case for all three processes shown in Fig.\ref{fig:class1}, where only one diagram enters. 
As a result, the whole amplitude within factorization contains only $\langle O_1\rangle_F$:
\begin{equation}
\langle D_s^- \pi^+ |{\cal H}_{eff} | B_s\rangle = \frac{G_F}{\sqrt{2}} V_{ud} V_{cb}^* \Big(
\ a_1 \ \langle O_1\rangle_F + C_2(m_b) \times nonfact \ 
\Big)
\quad \equiv \  T ,
\label{amplitudeT}
\end{equation}
where $
a_1 = C_1(\mu ) + \frac{C_2(\mu)}{N_c}.
$
Since $C_1 =1$ and $C_2=0$ at the fundamental (electroweak) scale, $C_2(m_b)$ is small and therefore the non-factorizable term is expected to be relatively  small. On the right hand side, we follow the notation of Ref. \cite{Fleischer:2010ca}
and refer to this colored-allowed amplitude as $T$. In contrast, $B^0\to D^-\pi^+$ (Fig.\ref{fig:class3}) has two diagrams: one of type $T$, dominant and similar to the case above, and the other, called $E$, the internal $W$ exchange, which is subdominant but contains relatively larger non-factorizable contributions. The presence of two diagrams corresponds to the two different contractions of the operators $O_1$ and $O_2$ for this process. Using a definition for the factorized matrix element $\langle D^-\pi^+|( \bar u c)|0\rangle\langle 0|(\bar b d) |B\rangle \equiv \langle O_2\rangle_F$, analogous to that of $\langle O_1\rangle_F$ in Eq.\eqref{O1F}, one can see that, for this process:
\[
\langle D^-\pi^+|O_1|B^0\rangle\  =  \langle O_1\rangle_F +
\frac{1}{N_c} \langle O_2 \rangle_F   + nonfact.
\]
and
\[
\langle D^-\pi^+|O_2|B^0\rangle\ = \langle O_2\rangle_F + \frac{1}{N_c} \langle O_1\rangle_F + nonfact .
\]

Thus the $\langle O_2\rangle_F$ terms add up to a second amplitude, called $E$ \cite{Fleischer:2010ca}:
\begin{equation}
E \ \equiv  \langle D^- \pi^+ |{\cal H}_{eff} | B^0\rangle_E \ \  \to \   
 \frac{G_F}{\sqrt{2}} V_{ud} V_{cb}^* \Big(\ a_2 \ \langle O_2\rangle_F + C_1(m_b) \times nonfact \ \Big) \ \  
 \equiv  E_{fact} + E_{nonfact}.
\label{amplitudeE}
\end{equation}
where $
a_2 = C_2(\mu ) + \frac{C_1(\mu)}{N_c}.
$
Here the non factorizable part is not suppressed relative to the term containing $\langle O_2\rangle_F$.  
Because of this uncertainty, the mode $B\to DK$ was preferred instead of $B\to D\pi$.

Now, the determination of $f_s/f_d$ using the method of Eq.\eqref{fsd} requires a theoretical expression for $B_s\to D_s\pi$, since there is no other independent measurement of this rate. Using $B\to DK$ instead of $B\to D\pi$ in the denominator has the advantage that the theoretical expression for it is similar and several factors tend to cancel, except for the form factors and uncertainties due to the non-factorizable  contributions. The non-factorizable contributions are  expected to be smaller in $B\to DK$ than in $B\to D\pi$, so $B\to DK$ was preferred even when the number of events of that mode is smaller. Using Eqs. \eqref{decayrate}, \eqref{AmplitudeBsDspi} and \eqref{AmplitudeBDK}, one obtains:
\begin{equation}
\frac{Br(B_s\to D_s^-\pi^+)}{Br(B^0\to D^- K^+)}
=\Phi_{sK} \frac{\tau_{B_s}}{\tau_{B^0}} \left|\frac{V_{ud}}{V_{us}} \right|^2\left(\frac{f_\pi}{f_K}\right)^2 \times {\cal N}_F^{(K)} \times {\cal N}_a.
\label{DspiDK}
\end{equation}
Here $\Phi_{sK} $ is a kinematic factor:
\[
\Phi_{sK} = \left( \frac{m_{Bs}^2 - m_{D_s}^2}{m_{B^0}^2 - m_{D^-}^2}\right)^2 
\left(\frac{m_{B^0}}{m_{B_s}}\right)^2 \frac{p_{CM}^{(s)}}{p_{CM}^{(K)}} \quad \simeq 1.0302,
\]
where $p_{CM}^{(s)} \simeq 2.3201$ GeV and $p_{CM}^{(K)} \simeq 2.27901$ GeV are the CM 3-momenta of the final particles in the corresponding decays. The factor ${\cal N}_F^{(K)}$ is 
the ratio of form factors at the corresponding kinematic points:
\[
{\cal N}_F^{(K)} = \left( \frac{f_0^{(s)}(m_\pi^2)}{f_0^{(d)}(m_K^2)}\right)^2 ,
\]
 while  ${\cal N}_a$ is the ratio of non-factorizable corrections. ${\cal N}_F^{(K)}$ and ${\cal N}_a$ are the main sources of theoretical uncertainty in Eq.\eqref{DspiDK}.

Now, using $Br(B^0\to D^-\pi^+)$ instead of $Br(B^0\to D^-K^+)$ in the denominator would presumably introduce larger uncertainties from non-factorizable contributions. We could avoid them if we use the experimental result of $B^0\to D^-\pi^+$ instead, which will be sensitive to the experimental error only, but then we will need the full theoretical estimate of the numerator, which includes the absolute value of the form factor, not just a ratio of form factors:
\begin{equation}
\frac{Br(B_s\to D_s^-\pi^+)}{Br(B^0\to D^- \pi^+)}
=\frac{ G_F^2 |V_{cb}V_{ud}|^2 a_1^2 f_\pi^2 \left[f_0^{(s)}(m_\pi^2)\right]^2 (m_{B_s}^2-m_{D_s}^2)^2 p_{CM}^{(s)}/(16\pi m_{B_s}^2 \Gamma_{B_s})}{Br(B^0 \to D^-\pi^+)_{exp}} .
\label{BDpi_ratio}
\end{equation}
Currently $Br(B^0 \to D^-\pi^+)_{exp} = (2.51 \pm 0.08)\times 10^{-3}$ \cite{ParticleDataGroup:2022pth}, i.e. it is known with a $3.2\%$ uncertainty.
Possibly Belle II may provide a measurement with even better precision in the near future. However the theoretical uncertainty in the numerator of Eq. \eqref{BDpi_ratio} is presumably larger.

Instead, let us then reevaluate the theoretical uncertainties using $Br(B^0\to D^-\pi^+)$. We will again have a cancellation of form factors (possibly better than in Eq.\eqref{DspiDK} by being both evaluated at the same $q^2$), but we need to assess the $E$ amplitude. 
Compared to Eq.\eqref{DspiDK}, in this case we have:
\begin{equation}
\frac{Br(B_s\to D_s^-\pi^+)}{Br(B^0\to D^- \pi^+)}
= \Phi_{s\pi} \frac{\tau_{B_s}}{\tau_{B^0}} \times {\cal N}_F^{(\pi)}\times {\cal N}_E ,
\label{DspiDpi}
\end{equation}
with the kinematic factor $\Phi_{s\pi} \simeq 1.0179$ (using now  $p_{CM}^{(\pi)} \simeq 2306.41$ MeV for $B\to D\pi$), 
while the ratio of lifetimes \cite{ParticleDataGroup:2022pth} 
$\tau_{B_s}/\tau_{B^0} = 1.0007 \pm 0.0042$ is known to a precision better than $0.5\%$.
The ratio of form factors is now:
\[
{\cal N}_F^{(\pi)} = \left( \frac{f_0^{(s)}(m_\pi^2)}{f_0^{(d)}(m_\pi^2)}\right)^2 ,
\]
and ${\cal N}_E$ is the correction to include the internal $W$-exchange diagram in $B^0\to D^-\pi^+$ (see Fig. \ref{fig:class3}), namely 
\[
{\cal N}_E = \left| \frac{T}{T + E}\right|^2,
\]
where $T$ and $E$ are the  color-allowed and 
$W$-exchange amplitudes respectively [Eqs.\eqref{amplitudeT} and \eqref{amplitudeE}] for the decay $B^0\to D^-\pi^+$. 

For the estimate of ${\cal N}_E$ we will use two theoretical results, which actually differ and therefore give us a caveat about the correct estimate of this factor.
Let us first consider Ref.~\cite{Wu:1996he},
which uses perturbative QCD including Sudakov factors to estimate the amplitudes of two-body non-leptonic $B\to D^{(*)}\pi(\rho)$ decays.
According to these authors, $E$ has comparable factorized and non-factorizable contributions, but both are much smaller than $T$ and mainly imaginary with respect to $T$:  in a normalization relative to $T\equiv 1$, they find $E_{fact}= -0.018 + 0.057\, i$, $E_{nonfact} = -0.0011+0.020\,  i$, which they add up to $E= -0.019 + 0.077\, i$.
Therefore, the
estimate of ${\cal N}_E$ that includes 
only factorizable terms would be 
$
{\cal N}_E^{(fact)} = 1/((1-0.018)^2 + (0.057)^2) \simeq 1.034
$,
while if the non-factorizable estimates are included the result extracted from Ref.~\cite{Wu:1996he}  is
\[
{\cal N}_E = \frac{1}{(1-0.019)^2 + (0.077)^2} \simeq 1.033\ .
\]
We could then use ${\cal N}_E \simeq 1.033$, where the uncertainty due to neglecting the non-factorizable contributions would be well below $1\%$ if those contributions were anything close to the estimated $E_{nonfact}$ \cite{Wu:1996he},
and thus negligible compared to other uncertainties in Eq.\eqref{DspiDpi}.

Alternatively,  let us consider Ref. \cite{Beneke:2000ry} for the estimation of  ${\cal N}_E$. Here the authors
study in systematic detail the different corrections for the factorization hypothesis, including those from the internal $W$-exchange amplitude $E$ (which they call ``annihilation amplitude'' $A$), as well as non-factorizable contributions to the dominant amplitude $T$. For the first correction they obtain $E/T_{lead} \simeq 0.04$ and the latter correction $T_{non.fact}/T_{lead} \simeq -0.03$. Considering all these estimates from Ref.~\cite{Beneke:2000ry}, we obtain 
\[
{\cal N}_E =\frac{1-0.03}{1-0.03+0.04} \simeq 0.97 ,
\]
while disregarding $T_{non.fact}$ the value changes in only $1\%$.
However, the discrepancy in this factor extracted from Refs.~\cite{Beneke:2000ry} and \cite{Wu:1996he} is larger. Taking both results as a range with a central value and an uncertainty, we get:
\[
{\cal N}_E \simeq 1.00 \pm 0.03
\]
i.e. a $3\%$ uncertainty.
Therefore, considering  this estimate on ${\cal N}_E$, still the main source of uncertainty for the ratio of Eq.\eqref{DspiDpi}
would come from the ratio of form factors, ${\cal N}_F^{(\pi)}$ which, according to Ref.\cite{LHCb:2021qbv} (see also \cite{Bordone:2019guc, Bordone:2020gao}) it is ${\cal N}_F^{(\pi)} = 1.000\pm 0.042$ (a $4\%$ uncertainty).
Consequently we would obtain the ratio
\begin{equation}
\frac{Br(B_s\to D_s^-\pi^+)}{Br(B^0\to D^- \pi^+)}
= 1.02 \pm 0.05 .
\label{DspiDpi1}
\end{equation}

It is interesting to notice that this result, which is a theoretical estimate of the ratio of branching ratios, slightly differs from the fit from data obtained by LHCb in Ref. \cite{LHCb:2021qbv}, which is $1.18 \pm 0.04$, i.e. a discrepancy in the central values by $2.5\, \sigma$, if we consider $\sigma$ as the quadrature combination of the errors in both results.
In order to find out the origin of the discrepancy, we cannot analyze the fit by LHCb, but we can examine the sources of uncertainty of our estimate shown in Eq. \eqref{DspiDpi1}. 

As seen in Eq. \eqref{DspiDpi}, there are only two possible sources of uncertainty at the level of a few percent or more: the form factor ratio ${\cal N}_F^{(\pi)}$, which is estimated to be known to $4\%$ as cited above, 
and the internal $W$ exchange correction ${\cal N}_E$ which,
according  to Refs.~\cite{Beneke:2000ry} and \cite{Wu:1996he}, should contain an uncertainty around $3\%$.
However, we should be cautious here: there are other estimates that differ from this result.
For example, ${\cal N}_E \simeq 0.966 \pm 0.05$ according to Ref. \cite{Fleischer:2010ca}. Notice that this value is also below unity, which means that the real part of $E$ should be positive instead of negative as in Ref.\, \cite{Wu:1996he}. But then again, the same Ref. \cite{Fleischer:2010ca} quotes a previous CDF result \cite{CDF:2008gud} that leads to a value ${\cal N}_E \sim 1.07 \pm 0.03$ (stat), i.e larger than unity.
In summary, there is still an uncertainty in ${\cal N}_E$ that needs to be resolved. Just considering the range of values ${\cal N}_E \sim 0.966$ \cite{Fleischer:2010ca} and $1.033$  \cite{Wu:1996he}, these differ in near 7$\%$. 

Now, if we take this more conservative uncertainty of 7\%,
the theoretical ratio of Eq.\eqref{DspiDpi1} would be known to that level of uncertainty. 
Then, measurements of $B_s\to D_s\pi^+$ and $B^0\to D^- \pi^+$ events at LHCb analogous to that of Eq.\eqref{fsd} could be used to extract the ratio of fragmentation fractions $f_s/f_d$, namely: 
\begin{equation}
 \frac{f_s}{f_d} \times \frac{Br(B_s \to D_s^-\pi^+)}{Br(B^0\to D^-\pi^+)} \ =\  \frac{n_{cor}(B_s\to D_s^-\pi^+)}{n_{cor}(B^0\to D^-\pi^+) },
 \label{fsDpi}
\end{equation}
where $n_{cor}$ are the efficiency-corrected experimental number of events just as in Eq.\eqref{fsd}. One could then extract the value for $f_s/f_d$ with a precision up to $7\%$. The actual values of $n_{cor}$ should be determined by experiment, which is of course beyond the scope of our work.

\section{On $f_c/f_d$ and $B_c^+\to J/\psi \, \pi^+$}

As we saw at the end of Section II, there is a tension between the theoretical predictions of $Br(B_c\to B_s\pi)$ and $Br(B_c\to J/\psi\, \pi)$ if one tries to relate these processes with $f_c/f_s$, $f_c/f_d$ and the experimental data.

LHCb \cite{LHCb:2013xlg} determined $Br(B_c\to B_s\pi)$ by measuring the quantity we reproduce here in Eq. \eqref{BcBs}, and by using $f_c/f_s$ as additional input. This input was in turn estimated using a previous result of $f_s/f_d$ and a value of $f_c/f_d$ obtained from their measurement shown in Eq.\eqref{fcfd} together with a broad range of theoretical estimates for 
$
Br(B_c\to J/\psi\,\pi) \sim 0.06\% \to 0.18\%
$
\cite{Ivanov:2006ni}. The resulting branching ratio $Br(B_c\to B_s\pi)\sim 5\% \to 15\%$, in spite of this broad range, is in slight tension with the direct theoretical estimates of $Br(B_c\to B_s\pi)$ shown in Table I and  Eq.\eqref{branchBcBs}:
in order to have consistency, one should expect 
$Br(B_c\to J/\psi \, \pi)$ to be lower than the aforementioned range.

Let us then  derive the values of $Br(B_c\to J/\psi \, \pi)$ that would be consistent with the theoretical estimates for $Br(B_c\to B_s \pi)$ given in Section II. 
To do that, first we take  $f_c/f_s$ in Eq.\eqref{fcfsnew}, 
which was derived using the theoretical estimates for $Br(B_c\to B_s \pi)$,  
and take $f_s/f_d = 0.2390\pm 0.0076$ obtained by LHCb at 7 TeV \cite{LHCb:2021qbv}. One then finds the values for $f_c/f_d$:
\begin{equation}
    \frac{f_c}{f_d} = \frac{f_c}{f_s}\times \frac{f_s}{f_d} = (1.925 \pm 0.291 \pm 0.068 \pm 0.061)\times 10^{-2}  a_1^{-2},
\end{equation}
where the first uncertainty comes from the LHCb measurement of Eq.\eqref{BcBs}, which is near $15\%$ and by far the largest, the second is near $3.5\%$ and comes from the lattice estimate of the $B_c\to B_s$ form factor shown in Eq.\eqref{eq:f02}, and the third is near $3.2\%$ and comes from the uncertainty in $f_s/f_d$ according to Ref.\cite{LHCb:2021qbv}.  

Now, using the LHCb measurement \cite{LHCb:2012ihf} shown in Eq.\eqref{fcfd} and the current value of  $Br(B^+\to J/\psi \, K^+) = (1.020\pm 0.019)\times 10^{-3}$ \cite{ParticleDataGroup:2022pth}, we get:
\begin{equation}
Br(B_c^+\to J/\psi \pi^+) = (3.60\pm 0.61 \pm 0.54 \pm 0.18) \times 10^{-4}\, a_1^2,
\label{BrBcJ}
\end{equation}
where the first uncertainty (near $17\%$) comes from the LHCb measurement shown in Eq.\eqref{fcfd}, the second uncertainty (near $15\%$) is from the LHCb measurement shown in Eq.\eqref{BcBs}, and the last uncertainty (near $5\%$) is the combination of the uncertainties from the branching ratio of $B^+\to J/\psi K^+$, the form factor of Eq.\eqref{eq:f02}, and $f_s/f_d$ shown above.
The combined uncertainty is near 23\%. Our result for $Br(B_c^+\to J/\psi \pi^+)$ in Eq.\eqref{BrBcJ} 
contrasts with the theoretical estimates \cite{Ivanov:2006ni}
used by LHCb \cite{LHCb:2013xlg} to extract $Br(B_c^+\to B_s\pi^+)$, namely:
\begin{equation}
Br(B_c\to J/\psi \pi) = (6\to 18)\times 10^{-4} .
\label{BJLHCb}
\end{equation}
Using $a_1 =1.14$ (the accepted coefficient $a_1$ for $b\to c$ transitions), our result in Eq.\eqref{BrBcJ} with its 23\% uncertainty is still $3\, \sigma$ below the lowest value in Eq. \eqref{BJLHCb}. 
Moreover, other theoretical models 
(see Table \ref{tab:brBcJpsipi})
predict values for  $Br(B_c^+\to J/\psi\, \pi^+)$ in the range
\hbox{$(3.8 \to 13)\times 10^{-4} a_1^{2}$}, and in particular the lowest two values in Table \ref{tab:brBcJpsipi} differ in less than $2\sigma$ from the value we calculated in Eq. \eqref{BrBcJ}.
While there are still large uncertainties in the estimation of the $B_c$ branching ratios and on the ratio of fragmentation functions, these latter comparisons tend to indicate that our results are approaching a consistent set of values.

\renewcommand{\arraystretch}{1.1}%
\begin{table}[h]
\footnotesize
    \centering
\begin{tabular}{l|c}
Model
  & $B_c^+\to J/\psi \pi^+$\\
\hline\hline
KLO \cite{Kiselev:1999sc,Kiselev:2002vz}
  & $9.8\times 10^{-4}\,a_1^2$\\
EFG \cite{Ebert:2003cn}
  & $4.68\times 10^{-4}\,a_1^2$\\
AMV \cite{AbdEl-Hady:1999jux}
  & $8.5\times 10^{-4}\,a_1^2$\\
CC \cite{Chang:1992pt}
  & $12.0\times 10^{-4}\,a_1^2$\\
CD \cite{Colangelo:1999zn}
  & $10.3\times 10^{-4} \,a_1^2$\\
AKNT \cite{Anisimov:1998xv}
  & $7.88\times 10^{-4}\,a_1^2$\\
LC \cite{Liu:1997hr} 
  & $7.06\times 10^{-4}\,a_1^2$\\
IKS \cite{Ivanov:2006ni}
  & $13.2\times 10^{-4} \,a_1^2$\\
NDKB \cite{Nayak:2022qaq}
  & $3.83\times 10^{-4}\,a_1^2$\\  
\hline
Our result [Eq. \ref{BrBcJ}
  & $3.60\times 10^{-4}\,a_1^2$\\  
\hline
\hline
\end{tabular}    
    \caption{Prediction for the branching ratio $Br(B_c^+\to J/\psi\pi^+$) from different models, and our estimate of Eq. \ref{BrBcJ}. The coefficient $a_1 = 1.14$ has been extracted for comparison between the different results.}
    \label{tab:brBcJpsipi}
\normalsize
\end{table}
\renewcommand{\arraystretch}{1.0}%

\section{Conclusions}

We have studied the consistency of current determinations and estimates of the fragmentation fractions $f_c$, $f_s$ and $f_d$, 
which are the probabilities that a $b$ quark produced at the LHCb hadronizes
into a $B_c$, $B_s$ and $B_d$ meson, respectively, together with the predictions of the branching ratios of these mesons into the specific two-body non-leptonic modes that are measured nowadays by LHCb.  We compare the theoretical estimates of the branching ratios obtained by different authors using different methods, and in our own estimates we use Lattice results for the corresponding form factors which enter in the theoretical expressions of these decays within the factorization approximation. In the analysis we include the uncertainties due to non-factorizable contributions as well. 

In particular, $f_c$ is estimated from the theoretical prediction of $B_c^+ \to B_s \pi^+$ based on lattice results, and  measurements by LHCb. 
This decay is a remarkable case, where the heavy quark is a spectator of the weak interaction
process, unlike the much studied decays of heavy hadrons where it is the heavy quark that goes through the weak transition. Current theoretical estimates of $Br(B_c^+\to B_s\pi^+)$ range from $0.01$ to $0.05$. Our estimate is $Br(B_c^+\to B_s\pi^+)\sim 0.03$, based on Lattice results, with an uncertainty of about 4\%, in which case $f_c/f_s\sim 0.08$ could be determined from the LHCb measurement with 15\% uncertainty.

Concerning the fraction  $f_s/f_d$, we study the possibility obtaining better results by measuring the ratio $Br(B_s \to D_s \pi)/Br(D^0 \to D \pi)$ instead of the proposed ratio
$Br(B_s \to D_s \pi)/Br(D^0 \to D K)$. The latter was proposed because $B\to D\pi$ contains non factorizable contributions, thus inducing possibly large uncertainties. However, $B\to D K$ has the disadvantage of being a more suppressed mode, so data is more limited. We therefore investigated the effect of non factorizable contributions in $B\to D\pi$ to see whether it could compensate the uncertainties by being a more abundant mode. In order to estimate $Br(B^0 \to D \pi)$ we have discussed various methods, such as
using the experimental value from PDG \cite{ParticleDataGroup:2022pth}, PQCD factorization predictions \cite{Beneke:2000ry,Wu:1996he} and the various fittings \cite{Fleischer:2010ca,CDF:2008gud}.
We find that the measurements of $B_s \to D_s\pi$ and $B^0\to D\pi$ at LHCb can then provide $f_s/f_d$ up to 7\% uncertainty, due to the corresponding non-factorizable contributions.

In summary, we predict $Br(B_c\to B_s\pi) \simeq 0.04$ (using $a_1 =1.2$ for this $c\to s$ transition), within 3.5\% uncertainty from the Lattice results for the form factor, which then gives $f_c/f_s \simeq 0.056$ with 16\% uncertainty using the LHCb measurements of this decay. 
Concerning $f_s/f_d$, we propose to use again the data on $B_s\to D_s\pi$ and $B\to D\pi$ considering our estimate of $Br(B_s\to D_s\pi)/Br(B^0\to D^-\pi^+) \simeq 1.02 \pm 0.05$, where the uncertainties include those from neglecting non factorizable contributions.
Finally, we obtain the ratio $f_c/f_d\simeq 0.014$ with a 16\% error based on $f_c/f_s$ and current $f_s/f_d$ results, and the estimate  $Br(B_c\to J/\psi \pi)\sim 4 \times 10^{-4}$ which is consistent with $Br(B_c\to B_s\pi)$ and the current data on the fragmentation fractions, with an uncertainty of 23\%. This consistent value is at the lower end of the current theoretical estimates, which still span a rather broad range, from $0.04\%$ to $0.17\%$.

\section*{Acknowledgements}
We thank Matthias Neubert for valuable conversations. We acknowledge support from FONDECYT (Chile) Grant No.\ 1210131 and ANID (Chile) PIA/APOYO AFB 180002, by NRF (Korea) grants NRF-2021R1A4A2001897 and NRF-2022R1I1A1A01055643,
and by ANID (Chile) FONDECYT Iniciaci\'on Grant No. 11230879 and ANID REC Convocatoria Nacional Subvenci\'on a Instalaci\'on en la Academia Convocatoria A\~no 2020, PAI77200092.

\bibliographystyle{unsrt}
\bibliography{bib}

\end{document}